\documentclass[pra,aps,twocolumn,showpacs,superscriptaddress,floatfix,amsmath,amssymb,nofootinbib]{revtex4}
\usepackage{amsfonts}
\usepackage{graphicx} 
\newcommand{\beq}{\begin{equation}}
\newcommand{\eeq}{\end{equation}}

\newcommand{\beqa}{\begin{eqnarray}}
\newcommand{\eeqa}{\end{eqnarray}}
\def\ra{\rangle}
\def\la{\langle}
\begin{document} 
\title{Fast transitionless expansions of cold atoms in optical 
Gaussian beam traps}
\author{E. Torrontegui}
\affiliation{Departamento de Qu\'{\i}mica F\'{\i}sica, Universidad del Pa\'{\i}s Vasco - Euskal Herriko Unibertsitatea, 
Apdo. 644, Bilbao, Spain}

\author{Xi Chen}
\affiliation{Departamento de Qu\'{\i}mica F\'{\i}sica, Universidad del Pa\'{\i}s Vasco - Euskal Herriko Unibertsitatea, 
Apdo. 644, Bilbao, Spain}
\affiliation{Department of Physics, Shanghai University, 200444 Shanghai, P. R. China}

\author{M. Modugno}
\affiliation{Departamento de F\'{\i}sica Te\'orica e Historia de la Ciencia, Universidad del Pa\'{\i}s Vasco - Euskal Herriko Unibertsitatea, 
Apdo. 644, Bilbao, Spain}
\affiliation{IKERBASQUE, Basque Foundation for Science}

\author{A. Ruschhaupt}
\affiliation{Institut f\"ur Theoretische Physik, Leibniz
Universit\"{a}t Hannover, Appelstra$\beta$e 2, 30167 Hannover,
Germany}

\author{D. Gu\'ery-Odelin}
\affiliation{Laboratoire Collisions Agr\'egats R\'eactivit\'e, CNRS UMR 5589, IRSAMC, Universit\'e Paul Sabatier, 118 Route de Narbonne, 31062 Toulouse CEDEX 4, France}

\author{J. G. Muga}
\affiliation{Departamento de Qu\'{\i}mica F\'{\i}sica, Universidad del Pa\'{\i}s Vasco - Euskal Herriko Unibertsitatea, 
Apdo. 644, Bilbao, Spain}

\affiliation{Max Planck Institute for the Physics of Complex Systems, 
N\"othnitzer Str. 38, 01187 Dresden, Germany}
\begin {abstract}
We study fast expansions of cold atoms in a three-dimensional Gaussian-beam optical trap. 
Three different methods to avoid final motional excitation are compared: 
inverse engineering using Lewis-Riesenfeld invariants, which provides the best overall performance, a bang-bang approach, and a fast adiabatic 
approach. We analyze the excitation effect of anharmonic terms, radial-longitudinal coupling, and radial-frequency mismatch. In the inverse engineering approach these 
perturbations can be suppressed or mitigated 
by increasing the laser beam waist.    
\end{abstract}  	
\maketitle
\section{Introduction: driven expansions of cold atoms}
Driving the expansion of a Bose-Einstein condensate or more generally of a cold atom cloud in a controlled way, by implementing a designed time-dependence of the trap frequency, is a common and basic operation in a cold-atom laboratory. The expansion may be aimed at different 
goals, such as decreasing the temperature \cite{Phillips, Kett}, adjusting the density to avoid three body losses \cite{Hansch}, facilitating temperature and density measurements
\cite{Hansch}, or changing the size of the cloud for further manipulations \cite{Kino}. Expansions, isolated \cite{ChenET10} or as part of refrigeration cycles \cite{Salamon09}, are also important from a fundamental point of view to quantify the third principle of thermodynamics.\footnote{Compressions may also play a role in many experiments 
but their treatment is similar to expansions so we shall not deal with them here explicitly.}   
In general, changes of the confining trap will excite the 
state of motion of the atoms unless they are done very slowly or, in the usual quantum-mechanical sense of the word, ``adiabatically'', but   
slow-change processes are also prone to 
perturbations and decoherence, or impractical for performing many cycles.  
Engineering fast expansions without final excitation 
is thus receiving much attention recently both theoretical 
and experimental  \cite{Salamon09,Muga09,Ch10,Muga10,ChenET10,MN10,Nice10,Li10,MN11,Nice11,Nice11b,optimal_control,Wu11,Adol11,Adol11b,Chen11,nonHermitian,Onofrio}. 
Most theoretical treatments so far are for idealized one-dimensional (1D) systems, but the implementation requires a three-dimensional (3D) trap \cite{Nice10,Nice11,Nice11b}, in principle with anharmonicities and couplings among different directions.  

\begin{figure}[t]
\begin{center}
\includegraphics[height=1.5cm,angle=0]{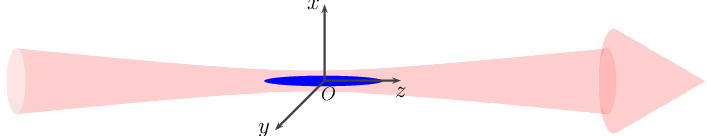}
\end{center}
\caption{\label{f0}
(Color online) Schematic view of the optical trap.
}
\end{figure}

In this paper we address these important aspects to implement the expansion in practice. Specifically we shall 
model a simple physical realization based on an elongated cigar-shaped optical dipole trap with cylindrical symmetry, see Fig. \ref{f0}. This trap is formed by a single laser which is red detuned with respect to an atomic transition to make the potential attractive (Section \ref{3D}), and is characterized in the harmonic approximation by longitudinal and radial frequencies. 
While magnetic traps allow for an independent control of longitudinal and radial frequencies \cite{Nice10,Nice11,Nice11b}, this is not the case for a simple laser trap
that therefore requires a special study.  
We assume that the time-dependence of the longitudinal frequency is engineered to avoid final excitations with a simple 1D harmonic theory (Section \ref{proto}) and analyze the final fidelity in the actual trap. Even though for full 3D-results we resort to a purely numerical calculation in Section \ref{full}, an understanding of the effects involved 
is achieved first by analyzing separately longitudinal and 
radial motions in Section \ref{lonrad}. 
Conclusions and open questions are drawn in the final Section \ref{dao}.
\section{The model\label{3D}}
%
%
%
%
%
The intensity profile of a Gaussian laser beam in the paraxial approximation is given by
\beq
\label{intensity}
I(r,z,t)=I_{0}(t)e^{-2r^2/w^2(z)}\frac{1}{1+z^2/z_{R}^{2}},
\eeq
where $r$ and $z$ are the radial and longitudinal coordinates respectively, and the variation of the spot size $w$
with $z$ is given by
$$
w(z)=w_0\sqrt{1+\left(\frac{z}{z_{R}}\right)^2},
$$
where $z_{R}=\pi w_{0}^{2}/\lambda$ is the Rayleigh range,  $w_0$ the waist, and $\lambda$ the laser wavelength. The paraxial approximation is valid for waists larger than about $2 \lambda/\pi$. 

If an atom is placed in this field with a detuning $\delta$ which is large 
with respect to the Rabi frequency $\Omega$ and the inverse of the lifetime of the excited state $\Gamma=\tau^{-1}$, but small with respect to the transition frequency, internal excitations and counter-rotating terms are negligible, and the potential that the ground state atoms feel due to the dipole force is proportional to the laser intensity, 
\beq
\label{pot}
V(r,z,t)\simeq\frac{\hbar\Omega^2}{4\delta}=\frac{\hbar\Gamma}{8}\frac{\Gamma}{\delta}\frac{I}{I_{sat}}, \quad |\delta|\gg\Omega, \Gamma, 
\eeq
and inversely proportional to $\delta$.  
$I_{sat}=\pi hc/(3\lambda^3\tau)$ is the saturation intensity.\footnote{Indeed other regimes could be used as well \cite{odt}. For larger detunings counter-rotating terms become important, but the dipole potential
is still proportional to the laser intensity. In the extreme limit of quasi-electrostatic traps the frequency of the trapping light is much smaller than the resonance frequency precluding the possibility of a potential sign change from attractive to repulsive; this 
change is in principle allowed by Eq. (\ref{pot}) by changing the sign of the detuning  across resonance, but see the discussion below.}   
Combining Eqs. (\ref{intensity})
and (\ref{pot}), 
and adding for convenience in later expressions 
the physically irrelevant term $V_0(t)$,
the potential takes finally the form
\beq
\label{V}
V(r,z,t)=-V_{0}(t)e^{-2r^2/w^2(z)}\frac{1}{1+z^2/z_{R}^{2}}
+V_0(t),
\eeq
where $V_0(t)=I_0(t)\hbar\Gamma^2/(8\delta I_{sat})$. 
The analysis carried out here is not restricted to a two-level atom. Indeed, the dipole force
is always proportional to the intensity for a multi-level atom but with a coefficient 
that depends on the level structure and the light polarization \cite{odt}. 
As an example of the magnitudes involved in practice,
for rubidium-87 atoms, a Gaussian beam linearly polarized of waist 8 $\mu$m, power $P=30$ W and wavelength 1060 nm provides a longitudinal trap frequency $\omega_{0z}/2\pi\simeq2500$ Hz. We shall use these or similar values in numerical examples below.  

In this work we shall assume small enough densities and times so that 
a one-particle, Schr\"odinger description for the atomic motion neglecting collisions 
is valid. 
Extensions to Tonks-Girardeau gases or condensates are possible as in \cite{Muga09,Ch10,Nice11}.
We shall also assume that the longitudinal axis lies horizontally and neglect the effect of gravity, either because it is artificially compensated, or because the radial confinement is tight.\footnote{This requires that the vertical shift of the trap center in the gravity field 
is small with respect to the waist, or $g\ll w_0\omega_R^2$, where $\omega_R$ is the radial angular frequency, and $g$ the acceleration due of gravity.}          
%
%
%
%
%
%
%
%
%
%
%
%
%
%
%

To solve the time dependent Schr\"odinger equation associated with the potential 
in Eq. (\ref{V}), 
\beq
i\hbar\frac{\partial\Psi}{\partial t}=-\frac{\hbar^2}{2m}\nabla^2\Psi+V\Psi, 
\eeq
%
we use cylindrical coordinates,    
$$ 
\nabla^2\equiv \frac{1}{r}\frac{\partial}{\partial r}\left(r \frac{\partial}{\partial r}\right)+\frac{1}{r^2}\frac{\partial^2}{\partial \phi^2}+\frac{\partial^2}{\partial z^2}.
$$
With the standard separation of variables $\Psi={\cal{F}}(r,z,t)\Phi(\phi)$, we find 
$$
-\frac{\hbar^2}{2m}\left[\frac{1}{r}\left(\frac{{\cal{F}}_r}{{\cal{F}}}+r
\frac{{\cal{F}}_{rr}}{{\cal{F}}}\right)+\frac{1}{r^2}\frac{\Phi_{\phi\phi}}{\Phi}+\frac{{\cal{F}}_{zz}}{{\cal{F}}}\right]+V=i\hbar\frac{{\cal{F}}_{t}}{{\cal{F}}},  
$$
(the simple or double subscripts denote first or second derivatives)  
whereas the equation for $\Phi$ reads 
\beq
\label{angu}
\frac{\Phi_{\phi\phi}}{\Phi}=-\nu^2 \quad \nu\in \mathbb{Z}, 
\eeq
where $\nu$ is a magnetic quantum number determining the conserved angular momentum 
component along the $z$ direction. This is solved by 
$e^{i\nu\phi}$ so we shall concentrate on ${\cal{F}}(r,z,t)$.   
Defining now $\tilde\Psi(r,z,t)=\sqrt{r}{\cal{F}}(r,z,t)$ we get
\beq
\label{sch}
i\hbar\frac{\partial\tilde\Psi}{\partial t}=-\frac{\hbar^2}{2m}\tilde\nabla^2\tilde\Psi+\tilde V\tilde\Psi,
\eeq
where
\beqa
\tilde\nabla^2&\equiv&\frac{\partial^2}{\partial r^2}+\frac{\partial^2}{\partial z^2}, \label{c} \\
\tilde V&\equiv&\frac{\hbar^2}{2m}\left(\frac{\nu^2-1/4}{r^2}\right)+V(r,z,t),
\label{p}
\eeqa
and $V(r,z,t)$ is given by Eq. (\ref{V}).
The task is now to design $V_0(t)$ so as to achieve a   
fast expansion without final excitation.    
\section{Expansion protocols\label{proto}}
In this section we briefly present three expansion protocols for a one-dimensional harmonic trap. They are chosen 
mostly because they have been realized experimentally
\cite{Phillips,Nice10,Nice11,Vogels}. Their simplicity will help us to identify relevant physical phenomena that will affect also other protocols.
No claim of completeness or global optimization is made, and 
there is certainly room for investigating other protocols. 

The objective is to expand the trap from an initial frequency $\omega_0/2\pi$ to a final frequency $\omega_f/2\pi$ in a time $t_f$, 
without inducing any final excitation in the adiabatic or instantaneous basis for which the Hamiltonian is diagonal. Note that transient excitations are 
generally allowed.
%
%
%
%
%
\subsection{Inverse engineering with Lewis-Riesenfeld invariants \label{erm}}
Lewis-Riesenfeld invariants may be used to engineer  
efficient expansion protocols  as explained in \cite{Muga09,Ch10,ChenET10,Nice10}. We refer the reader to these 
works for the details and give here only the elements required for a practical application. For any harmonic oscillator expansion, and in fact for any potential  
with the structure  
\beq
\label{Vinv}
V(q,t)=\frac{m}{2}\omega^2(t)q^2+\frac{1}{b^2(t)} U\left(\frac{q}{b(t)}\right),
\eeq
where the function $U$ is arbitrary, 
there is a quadratic-in-momentum invariant operator of the form   
\beqa
\label{I}
{\cal{I}}(q,p,t)&=&\frac{1}{2m}[b^2p^2-m\dot b b (qp+pq) + m^2\dot{b}^2q^2]
\nonumber\\
&+&\frac{1}{2}m\omega_0^2\left(\frac{q}{b}\right)^2
+U\left(\frac{q}{b}\right).
\eeqa 
where $q$ and $p$ are generic position and momentum operators (particularized later for 
longitudinal or radial directions),
and $b=b(t)$ is a scaling function that satisfies the Ermakov equation 
\beq
\label{Erma}
\ddot{b}+\omega^2(t) b=\omega_0^2/b^3. 
\eeq
To inverse engineer the trap frequency, $b(t)$ is 
designed to satisfy the boundary conditions
that guarantee no final excitations and continuity of the trap frequency, 
\beqa
b(0)=1,\, \dot{b}(0)=0,\, \ddot{b}(0)=0,
\nonumber\\
b(t_f)=\gamma,\, \dot{b}(t_f)=0,\, \ddot{b}(t_f)=0, 
\label{boco}
\eeqa
where $\gamma = \sqrt{\omega_0/\omega_f}$.  
Finally $\omega(t)$ is deduced from Eq. (\ref{Erma}).
When these conditions are satisfied the $n$-th eigenstate of $H(0)=p^2/(2m)+V(q,0)$ 
evolves as an ``expanding mode'', which is the time-dependent $n$-th eigenvector
of the invariant times a ``Lewis-Riesenfeld'' phase factor $e^{i\alpha_n}$, where   
$\alpha_n(t)=-(n+1/2)\omega_0\int_0^t dt'/b^2$, until it becomes an  
$n$-th eigenvector of the final Hamiltonian $H(t_f)=p^2/(2m)+V(q,t_f)$.  

Discontinuities in $\ddot{b}$ may in principle be allowed, but they amount to perform finite jumps of the trap frequency. This is an idealization but it can be approached up to technical limits. Discontinuities in $\dot{b}$ have more serious consequences, as they imply delta functions for $\ddot{b}$ and infinite trap frequencies at the jumps. Thus their approximate physical realization is even more difficult, but they are useful to find bounds for physical variables of interest as in \cite{ChenET10} or in Appendix \ref{apa}. There are many $b(t)$ that satisfy Eq. (\ref{boco}). A simple smooth function is a quintic polynomial with six coefficients,
%
%
\beq
b (t) =
6 \left(\gamma -1\right) s^5
-15 \left(\gamma-1\right) s^4 +10 \left(\gamma-1\right)s^3
+ 1,
\label{ieb}
\eeq
where $s:=t/t_f$. 
More sophisticated choices are possible  to 
minimize the process time, 
the average energy, or other variables with or without imposed constraints (``bounded control'') on the allowed trap frequencies \cite{Li10,transport3}.  

In principle this method allows for arbitrarily small values of $t_f$ but for very small  process times the potential becomes transitorily repulsive \cite{Ch10}. 
With a laser beam, a repulsive potential could be achieved 
by means of a positive (blue) detuning. However the validity of Eq. (\ref{V}) requires a large detuning, so care should be exercised when crossing
the resonance. 
In this work the numerical examples are restricted to positive trap frequencies
since the transient passage from attractive to a repulsive interaction could involve unwanted effects 
such as radiation pressure. We shall in any case point out  
in the figures the minimum value of $t_f$ for which the potential stays 
attractive for all times with the quintic $b(t)$.   
\subsection{Bang bang\label{bb}} 
Bang-bang methods are those with a  
stepwise constant behavior of some variable. The simplest one to avoid 
final excitations consists of 
using a constant intermediate trap frequency,
the geometric average of the initial and final frequencies,   
\beq
\omega^{bang}=\left\{
\begin{array}{ll} 
\omega_{0}, & t\leq 0
\\
(\omega_{0}\omega_{f})^{1/2}, &0<t<t_{f}^{bang}
\\
\omega_f, &t\geq t_{f}^{bang}=\pi/[2(\omega_{0}\omega_{f})^{1/2}] 
\end{array}\right..
\label{bang}
\eeq
during a fourth of the 
corresponding period.
Optimal bang-bang trajectories with two intermediate
frequencies and arbitrary final times are also possible \cite{Salamon09,Ch10}
but they are not considered here.  
To arrive at Eq. (\ref{bang}), one may use a classical argument 
\cite{Vogels} or 
proceed as follows:  
the solution of the Ermakov equation for $t>0$ assuming a constant 
intermediate angular frequency $\omega_1$ with initial boundary conditions $b(0)=1$, $\dot{b}(0)=0$ is 
\beq
b(t)=\sqrt{[(\omega_0^2-\omega_1^2)/\omega_1^2]\sin^2(\omega_1 t)+1}.
\eeq
Then, if we solve for $t_f$ and $\omega_1$ 
the two equations implied by the final boundary conditions 
\beq
b(t_f)=\gamma,\;\; \dot{b}(t_f)=0, 
\eeq
the values in Eq. (\ref{bang}) are found. 
\subsection{Fast adiabatic protocols} 
``Fast'' and ``adiabatic'' may appear as contradictory concepts. A fast adiabatic protocol seeks to perform a process as quickly as possible keeping it adiabatic at all times. For harmonic oscillator expansions the adiabaticity condition reads $\dot{\omega}/\omega^2\ll 1$, so making $\dot{\omega}/\omega^2$ constant from $t=0$ to $t_f$ \cite{Phillips}, and solving the resulting differential equation for $\omega(t)$, we get \cite{Ch10} 
\beq
\omega^{adi}(t)=\frac{\omega_{0}}{1-(\omega_{f}-\omega_{0})\frac{t}{t_f\omega_{f}}}.
\label{adiaba}
\eeq
\section{Longitudinal and radial motions\label{lonrad}} 
By expanding the potential in Eq. (\ref{V}) in a double Taylor series around $(z=0,r=0)$, 
%
\beqa
V(r,z,t)&\simeq&-V_0(t)\bigg(-\frac{2r^2}{w_{0}^{2}}-\frac{z^2}{z_{R}^{2}}
\nonumber \\ 
&+&\frac{2r^4}{w_{0}^{4}}+\frac{z^{4}}{z_{R}^{4}}+\frac{4r^2z^2}{w_{0}^{2}z_{R}^{2}}+\cdots\bigg),  
\label{pote}
\eeqa
we see that the first coupling term between radial and longitudinal motions is of fourth order, proportional to $r^2z^2$. If we could neglect this and higher order  
coupling terms, longitudinal and radial motions would be independent. The approximation that considers longitudinal and radial motions completely uncoupled is useful to analyze several effects separately and gain insight. Not only that, as we shall confirm later with full 3D calculations, this is also a good approximation 
numerically for low energy levels.   
%
%
%
%
\subsection{Longitudinal motion}
To study the motion in the longitudinal direction we consider first the full longitudinal Hamiltonian (putting $r=0$ in Eq. (\ref{V})) 
\beq
H(z,t)=-\frac{\hbar^2}{2m}\frac{\partial^2}{\partial z^2}-V_0(t)\left(\frac{1}{1+z^2/z_{R}^{2}}-1\right).
\label{longH}
\eeq
To characterize this potential in terms of a longitudinal frequency 
consider the harmonic approximation   
\beqa
\label{hz}
H_{har}(z,t)&=&-\frac{\hbar^2}{2m}\frac{\partial^2}{\partial z^2}+V_0(t)\frac{z^2}{z_{R}^2} \nonumber \\
&=&\frac{p_{z}^2}{2m}+\frac{m\omega_{z}^{2}(t)z^2}{2},
\eeqa
where $p_z=-i\hbar\partial/\partial z$ and
\beq
\label{oml}
\omega_{z}^2(t)=\frac{2V_0(t)}{mz_{R}^{2}}.
\eeq
We may now apply the methods described in Sec. \ref{proto}. 
We shall add a subscript $z$, for ``longitudinal'',
to the trap frequencies and take $q\to z$.    
For an imposed $\omega_{z}(t)$ and fixed waist and laser frequency, 
Eq. (\ref{oml}) determines the time dependence of the laser intensity. 

%
%
%
\begin{figure}[t]
\begin{center}
\includegraphics[height=5.cm,angle=0]{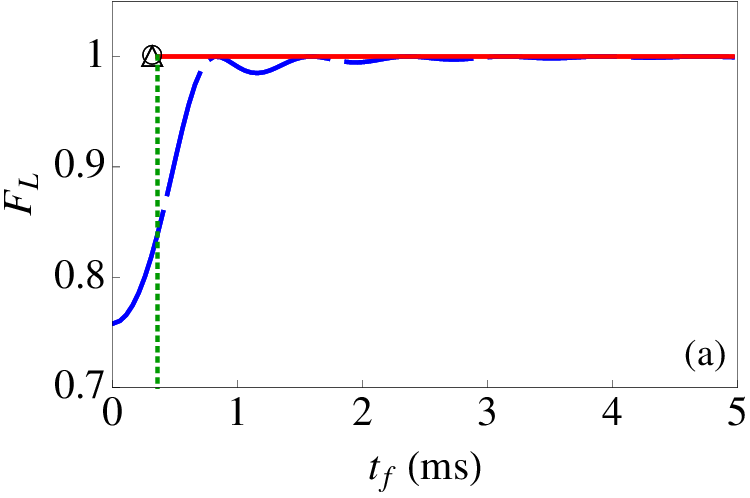}\vspace{0.1cm}
\includegraphics[height=5.cm,angle=0]{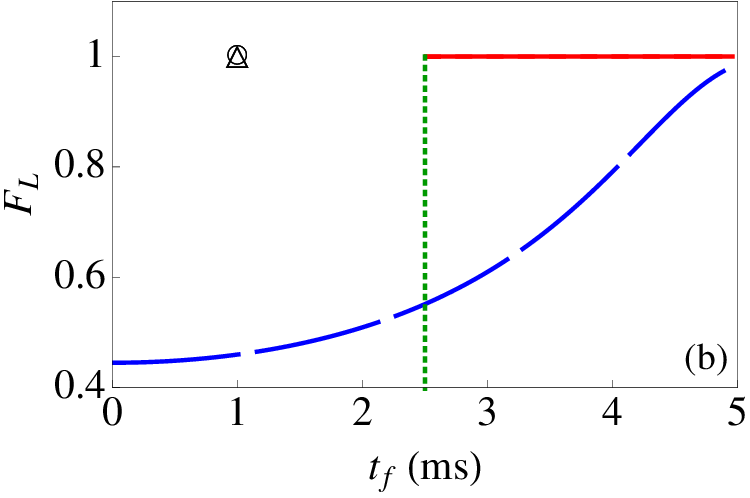}\
\end{center}
\caption{\label{f1}
(Color online) Longitudinal fidelity for Eq. (\ref{longH}) versus final time $t_f$  
for two different final frequencies:   
in (a) $\omega_{fz}/2\pi=250$ Hz;  in (b) $\omega_{fz}/2\pi=25$ Hz. 
In both figures 
$\omega_{0z}/2\pi=2500$ Hz,  
$\lambda=1060$ nm, and the initial state is the ground state. All curves have been calculated for two different waists, $w_0=10$ $\mu m$ and $w_0=3$ $\mu m$, but the results are indistinguishable.
$V_0(t)$ is chosen according to the three protocols:
inverse engineering, using Eq. (\ref{ieb}) in Eq. (\ref{Erma}) (solid red line);  
bang-bang (circles for $w_0=10$ $\mu m$
and triangles for $w_0=3$ $\mu m$); 
and fast adiabatic method (long-dashed blue line).    
The green vertical lines denote the minimum 
$t_f$ for which $V_0(t)$ remains positive for all $t$ in the inverse engineering  approach with a quintic $b(t)$.  
}
\end{figure}
%
\begin{figure}[t]
\begin{center}
\includegraphics[height=5.cm,angle=0]{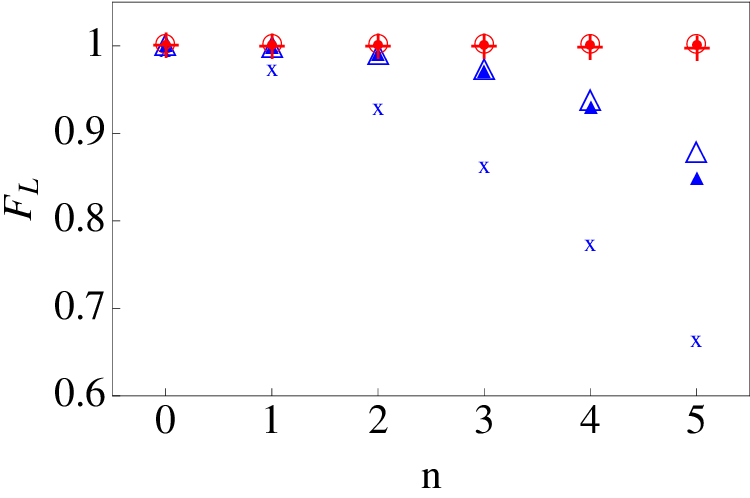}
\end{center}
\caption{\label{fnive}
(Color online) Longitudinal fidelity 
versus level number $n$. For $w_0=3$ $\mu m$ (blue symbols): 
fidelity $F_L=|\la Z_n(t_f)|U_z(t_f,0)|Z_n(0)\ra|$ with inverse engineering computed for Eq. (\ref{longH}) and a quintic $b(t)$ (empty triangles); 
first-order bound in Eq. (\ref{Fbound}) (x); second-order approximate fidelity computed with
Eqs. (\ref{Fper1}) and (\ref{Fper2}) for the quintic $b(t)$ (filled triangles). 
For a larger waist,        
$w_0=10$ $\mu m$ (red symbols), the corresponding values (empty circles, crosses, and filled circles), are very nearly one and indistinguishable. 
In all cases $\omega_{fz}/2\pi=25$ Hz,  $\omega_{0z}/2\pi=2500$ Hz,  
$\lambda=1060$ nm, and $t_f=2.5$ ms.   
}
\end{figure}
%
Figure \ref{f1} shows for the ground state, $n=0$, the ``longitudinal fidelity'' $F_L=|\la Z_n(t_f)|U_z(t_f,0)|Z_n(0)\ra|$, where $Z_n(z,0)$ and $Z_n(z,t_f)$
are the initial and final $n$-th eigenstates of the full longitudinal Hamiltonian, Eq. (\ref{longH}), for frequencies $\omega_{0z}$ and $\omega_{fz}$ respectively,
and $U_z(t_f,0)$ is the evolution operator with Eq. (\ref{longH}). 
As in all figures hereafter the
initial longitudinal  frequency of the trap is $\omega_{0z}/2\pi=2500$ Hz, and the wavelength is taken as $\lambda=1060$ nm, characteristic of  Neodymium-doped  lasers.  
The vertical green lines in this and the following figures 
correspond to the minimum $t_f$ value for which $\omega_z(t)$ is positive for all $t$ using the quintic $b(t)$. Of course this limit only applies to the 
inverse engineering approach. 
  

The three methods are compared for two different final frequencies,  
in (a) $\omega_{fz}/2\pi=250$ Hz and in (b) $\omega_{fz}/2\pi=25$ Hz,  
and two different waists, see the caption for details. Actually the effect of the waist change is negligible in the scale of these figures. 
Globally the inverse engineering method outperforms the others. 
The bang-bang approach provides a good fidelity, but only for a specific final time,  
and the fast adiabatic method fails in fact to be adiabatic at short times. This is very evident for the smaller final frequency in Fig. \ref{f1} (b):   
the opening of the trap is faster and the level spacings smaller than for 
the larger final frequency, so the state cannot follow an adiabatic behavior. 

For small $t_f$ and/or large $\gamma$ the transient excitation energy
during the inverse engineering protocol may be high, 
giving the atoms access to the anharmonic part of the potential.
This could lead to a decay of fidelity as $t_f$ decreases,
but the effect is not seen in the scale of Fig. \ref{f1}.    
The small effect of anharmonicity is enhanced by increasing the
vibrational number, see Fig. \ref{fnive}.   
It can be avoided by increasing the waist,   
which reduces the anharmonic terms, as demonstrated in Fig. \ref{fnive}.  
The Appendix \ref{apa} provides a perturbation theory
analysis of longitudinal 
anharmonicity. A first order approach gives analytical lower bounds
for the fidelity, 
and the possibility to maximize them by other choices of $b(t)$. More accurate results are found in second order, see Fig. \ref{fnive}.   

\subsection{Radial motion}
To study radial motion we define the radial Hamiltonian
by setting $z=0$ in Eq. (\ref{V}), 
and add the ``centrifugal term'' (an attractive one for $\nu=0$),
\beqa
H(r,t)&=&-\frac{\hbar^2}{2m}\frac{\partial^2}{\partial r^2}-V_0(t)\left(e^{-2r^2/w_{0}^{2}}-1\right)\nonumber\\
&+&\frac{\hbar^2}{2m}\left(\frac{\nu^2-1/4}{r^2}\right), \label{hr} 
\eeqa
In the harmonic approximation we take  
\beq
-V_0(t)\left(e^{-2r^2/w_{0}^{2}}-1\right)\sim\frac{m\omega_{R}^2(t)r^2}{2},
\label{happ}
\eeq
where  
\beq
\omega_{R}^{2}(t)\equiv\frac{4V_{0}(t)}{m w_{0}^{2}},
\label{omr}
\eeq
defines the radial frequency of the trap $\omega_{R}(t)/(2\pi)$.

We assume that $V_0(t)$ is set to satisfy the designed longitudinal expansion
according to Eq. (\ref{oml}). Substituting this into Eq. (\ref{omr})  
gives the relation between    
radial and longitudinal frequencies,  
\beq
\label{conexion}
\omega_{R}(t)=\frac{\sqrt{2} \pi w_{0}}{\lambda}\omega_{z}(t),
\eeq
which is key to understand the behavior of the radial wavefunction. 
The waist value $w_0=\lambda/(\sqrt{2}\pi)$ would make both frequencies equal but 
for such a small waist the paraxial approximation fails 
and the present theory could not be applied \cite{Nemoto}.

\begin{figure}[t]
\begin{center}
\includegraphics[height=5.cm,angle=0]{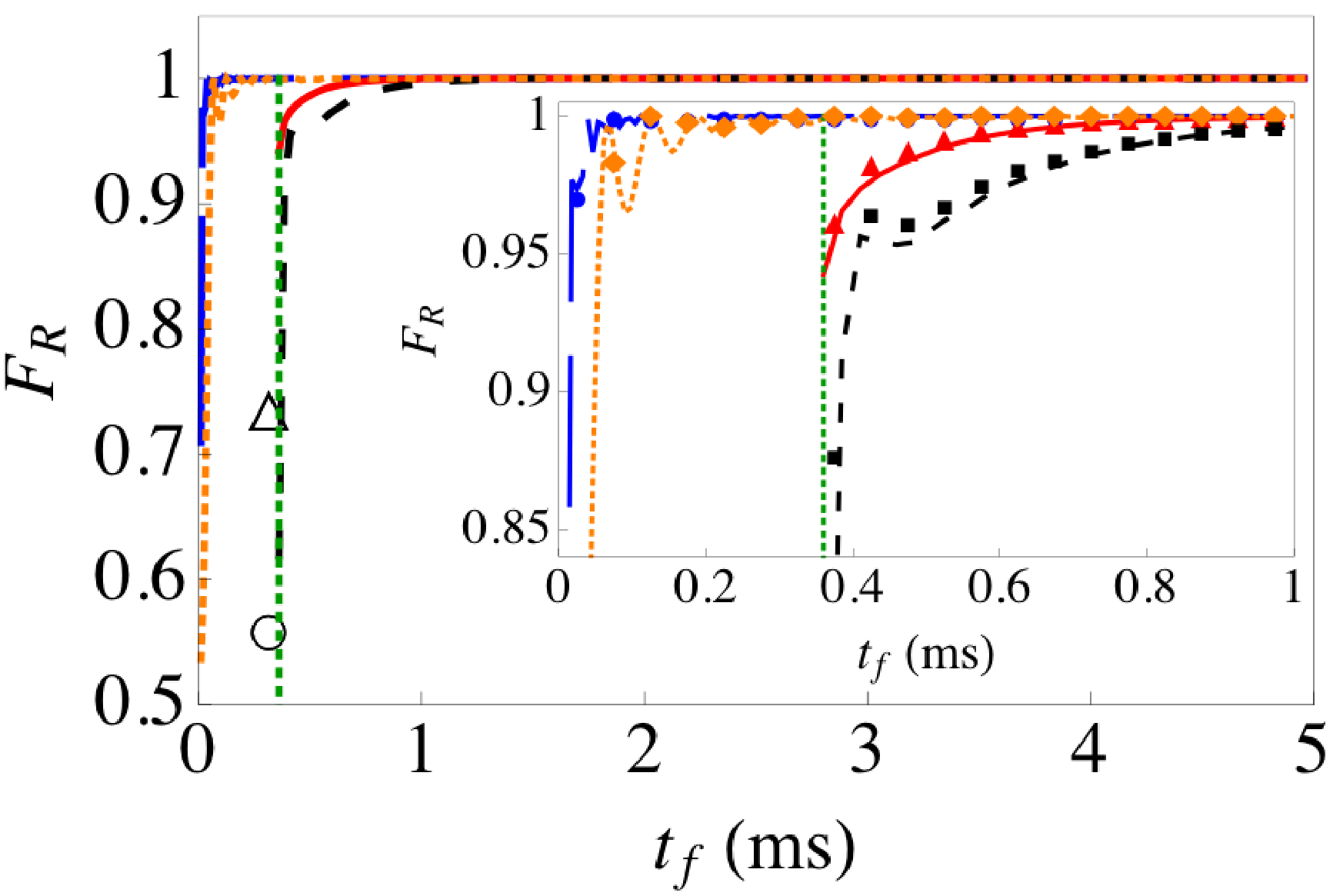}
\end{center}
\caption{\label{f2}(Color online) Radial fidelity for different final times.  
Parameters: $\nu=0$, $\omega_{0z}/2\pi=2500$ Hz, $\omega_{fz}/2\pi=250$ Hz, and 
$\lambda=1060$ nm.
Protocols: 
inverse engineering, quintic $b$, $w_0=10$ $\mu$m (solid red line); 
inverse engineering, quintic $b$, $w_0=3$ $\mu$m  (short-dashed black line);
bang bang, $w_0=10$ $\mu$m (empty circle); 
bang bang, $w_0=3$ $\mu$m (empty triangle); 
fast adiabatic, $w_0=10$ $\mu$m (long-dashed blue line);
fast adiabatic, $w_0=3$ $\mu$m  (dotted orange line).  
In the inset, corresponding (red) triangles, (black) squares, (blue) circles, and (orange) diamonds are the approximate fidelities from 
adiabatic perturbation theory using Eq. (\ref{overlap}). 
The green vertical line near $0.4$ ms is the minimal final time $t_f$ for which the quintic-$b$ inverse engineering protocol implies positive frequencies at all transient times.}
\end{figure}

Now we define the ground state ``radial fidelity'' as $F_R=|\la \phi_0(t_f)|U_R(t_f,0)| \phi_0(0)\ra|$, where 
$U_R(t_f,0)$ is the radial evolution operator for $H(r,t)$, and $\phi_0(r,0)$ and $\phi_0(r,t_f)$ are ground states of
the initial and final radial traps for Eq. (\ref{hr}), with $V_0(t)$ given by Eq. (\ref{oml}).
In Fig. \ref{f2} the radial fidelity is depicted for the three protocols explained above
and $\nu=0$. 
%
%
The bang-bang trajectory causes much excitation and gives a rather poor fidelity in the radial direction (see the big empty symbols). This is because, even though the ``correct'' transient radial frequency $\omega_{R}^{bang}(t)=(\sqrt{2}\pi w_0/\lambda)\omega_{z}^{bang}(t)$ is  
implemented (as the geometric 
average of initial and final radial frequencies), the time $t_f$ is adjusted for
the longitudinal, not for the radial frequency,  compare $t_{fz}^{bang}$ to $t_{fR}^{bang}:=\lambda t_{fz}^{bang}/(\sqrt{2}\pi w_0)$.
 
%
  
The behavior of the other two methods is much more robust 
as shown by their fidelities in Fig. \ref{f2}, where 
the inset amplifies the small-$t_f$ region.     
The fast adiabatic approach, blue long-dashed lines of Fig. \ref{f2}, gives an excellent radial fidelity. A detailed calculation shows that the 
adiabaticity condition for the radial Hamiltonian, see the Appendix \ref{apb}, takes, 
in spite of the centrifugal term, 
the same form as for the ordinary harmonic oscillator, namely $\dot{\omega}_R/\omega_R^2\ll 1$. 
If $\dot{\omega}_z/\omega_z^2$ is constant, the ratio $\dot{\omega}_R/\omega_R^2$ will be constant too, but smaller by a factor $\lambda/(\sqrt{2}\pi w_0)$, so that 
the radial dynamics will be in general more adiabatic.  
Indeed, 
%
comparing  blue long-dashed lines of Figs. \ref{f1} and \ref{f2}, 
we see that the radial fidelity is better than the longitudinal one, as the confinement is tighter radially than longitudinally, and the energy levels are more separated. 

\begin{figure}[t]
\begin{center}
\includegraphics[height=4.cm,angle=0]{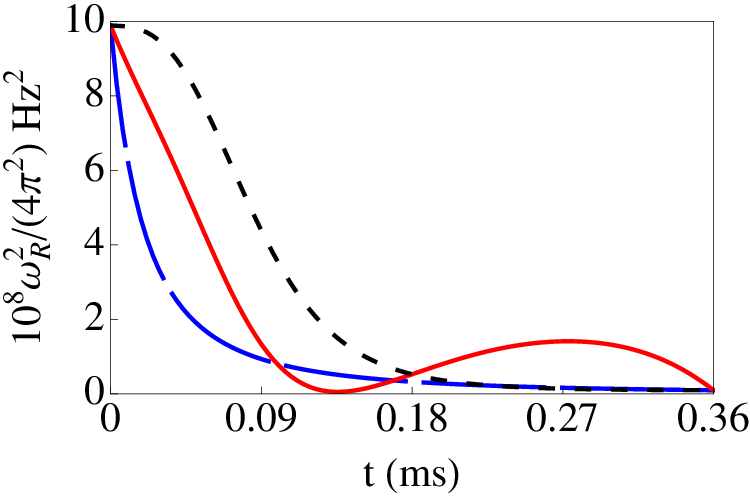}
\end{center}
\caption{\label{omRa}
(Color online) (Square of) radial frequency versus time:   
Eq. (\ref{orreal}) based on a quintic $b$ (red solid line); fast adiabatic protocol,  Eq. (\ref{conexion}) with Eq. (\ref{adiaba}) for $\omega_z$ (blue long dashed line);  Eq. (\ref{orinv}) (black short dashed line).  
Parameters: $\lambda=1060$ nm, $\omega_{0z}/2\pi=2500$ Hz, $\omega_{fz}/2\pi=250$ Hz, $w_0=3$ $\mu$m and $t_f=0.36$ ms.
}
\end{figure}

\begin{figure}[t]
\begin{center}
\includegraphics[height=5.cm,angle=0]{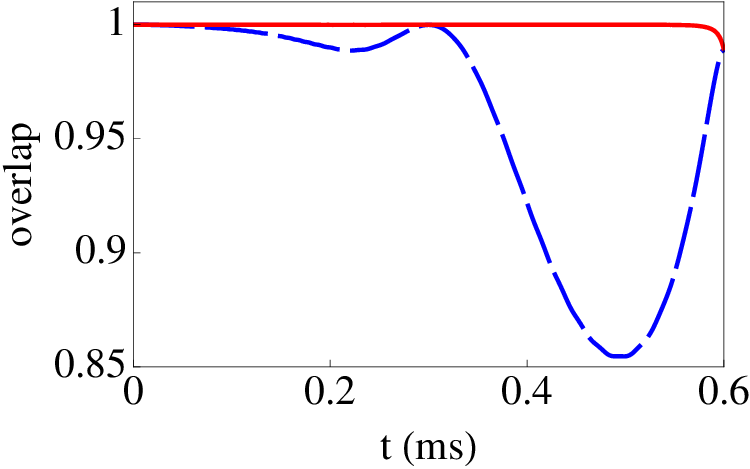}
\end{center}
\caption{\label{solap}
(Color online) (Modulus of) the overlap integral between the state evolving from the ground state with the radial Hamiltonian, using the actual radial frequency in Eq. (\ref{orreal}), 
and  
(a) the instantaneous eigenstate in harmonic approximation (red solid line), or 
(b) the expanding mode in harmonic approximation for the radial frequency in Eq. (\ref{orinv})
(dashed blue line).    
Parameters: $\nu=0$, $\omega_{0z}/2\pi=2500$ Hz,
$\lambda=1060$ nm,  
$\omega_{fz}/2\pi=250$ Hz, $w_0=3$ $\mu$m, $t_f=0.6$ ms.} 
\end{figure}

For the inverse engineering protocol  
with a quintic $b$, red solid and black short-dashed lines in Fig. \ref{f2}, the radial 
excitation is small, the fidelity being nearly one for $t_f>1$ ms in the scale of the figure. This could be surprising since the radial 
frequency does not behave according to an ideal inverse engineering  
dependence  
%
\beq
{\omega_{R}^{2}}_{inv}(t)=\frac{2\pi^2w_{0}^{2}}{\lambda^2}\frac{\omega_{0z}^{2}}{b^4}-\frac{\ddot b}{b},
\label{orinv}
\eeq
where we have used the Ermakov equation and  $\gamma=(\omega_{0z}/\omega_{fz})^{1/2}=(\omega_{0R}/\omega_{fR})^{1/2}$,
so the function $b$ is the same for the longitudinal and radial directions. Instead, the 
actual radial frequency varies according to Eq. (\ref{conexion}),
\beq
\omega_{R}^{2}(t)=\frac{2\pi^2w_{0}^{2}}{\lambda^2}\left(\frac{\omega_{0z}^2}{b^4}-\frac{\ddot b}{b}\right).
\label{orreal}
\eeq
%
The difference between Eqs. (\ref{orinv}) and (\ref{orreal}) 
may be quite significant, as the example in Fig. \ref{omRa} shows,  
so the radial state 
does not really follow the expanding modes corresponding to Eq. (\ref{orinv}). 
As a consequence a perturbative approach based on Eq. (\ref{orinv}) as the zeroth order fails
completely, even for qualitative guidance. (Further evidence is provided in Fig. \ref{solap}.)     
Instead, a valid perturbation theory analysis 
may be based on the zeroth order of the adiabatic modes, 
which are instantaneous eigenstates of the radial Hamiltonian.  
The physical reason for the relatively good radial behavior of the invariant-based 
inverse engineering protocol is thus the adiabaticity in that direction. When $t_f$ approaches
the critical lower value, $\omega_R(t)$ becomes too steep at small values
of $\omega_R$, see Fig. \ref{omRa}, and adiabaticity breaks down,
which explains the fidelity decrease there. To mitigate this problem,      
a larger waist may be used to increase $\omega_R(t)$ for a given $\omega_z(t)$, see Eq. (\ref{conexion}), making the radial process more adiabatic and improving  
the fidelity, as shown in Fig. \ref{f2}. Keep in mind 
that to implement a given $\omega_z(t)$ an increase of waist has to be compensated by an increase of laser intensity.
%

%
Inserting a wave function expansion in an adiabatic basis (instantaneous 
eigenstates) 
$\psi(r,t)=\sum_k a_k(t) \la r|\phi_k(t)\ra e^{-i\int_0^t E_k(t') dt'}$, 
into the radial Schr\"odinger equation provides a set of coupled differential equations. Integrating formally for $a_k(t)$ and 
assuming that in zeroth order  $a_k^{(0)}(t)=\delta_{0,k}$,     
we get in first order 
\cite{Schiff,Rice03}
\beqa
a_1^{(1)}(t)&=&-\int_0^t dt' \la \phi_1(t')|\dot{\phi}_0(t')\ra e^{-\frac{i}{\hbar}\int_0^{t'} dt''[E_0(t'')-E_1(t'')]}
\nonumber\\
&=&-\int_0^t dt'\frac{\dot{\omega}_R(t')}{2\omega_R(t')} e^{2i\int_0^{t'} dt'' \omega_R(t'')}, \label{overlap}
\eeqa
where the second line follows by applying, in addition, the harmonic approximation. 
All other first order amplitudes (for $k\geq 2$) are zero so the radial fidelity at 
$t_f$ can be calculated in this approximation as $F_R(t_f)=[1-|a^{(1)}_{1}(t_f)|^2]^{1/2}$.
This is in fact correct up to second order and it provides good agreement with the exact results as shown by the symbols in the inset of Fig. 
\ref{f2}. 
%
%
%
%
%
%
%
%
\section{Full 3D analysis\label{full}} 
%
%
%
%
\begin{figure}[t!]
\begin{center}
\includegraphics[height=5.5cm,angle=0]{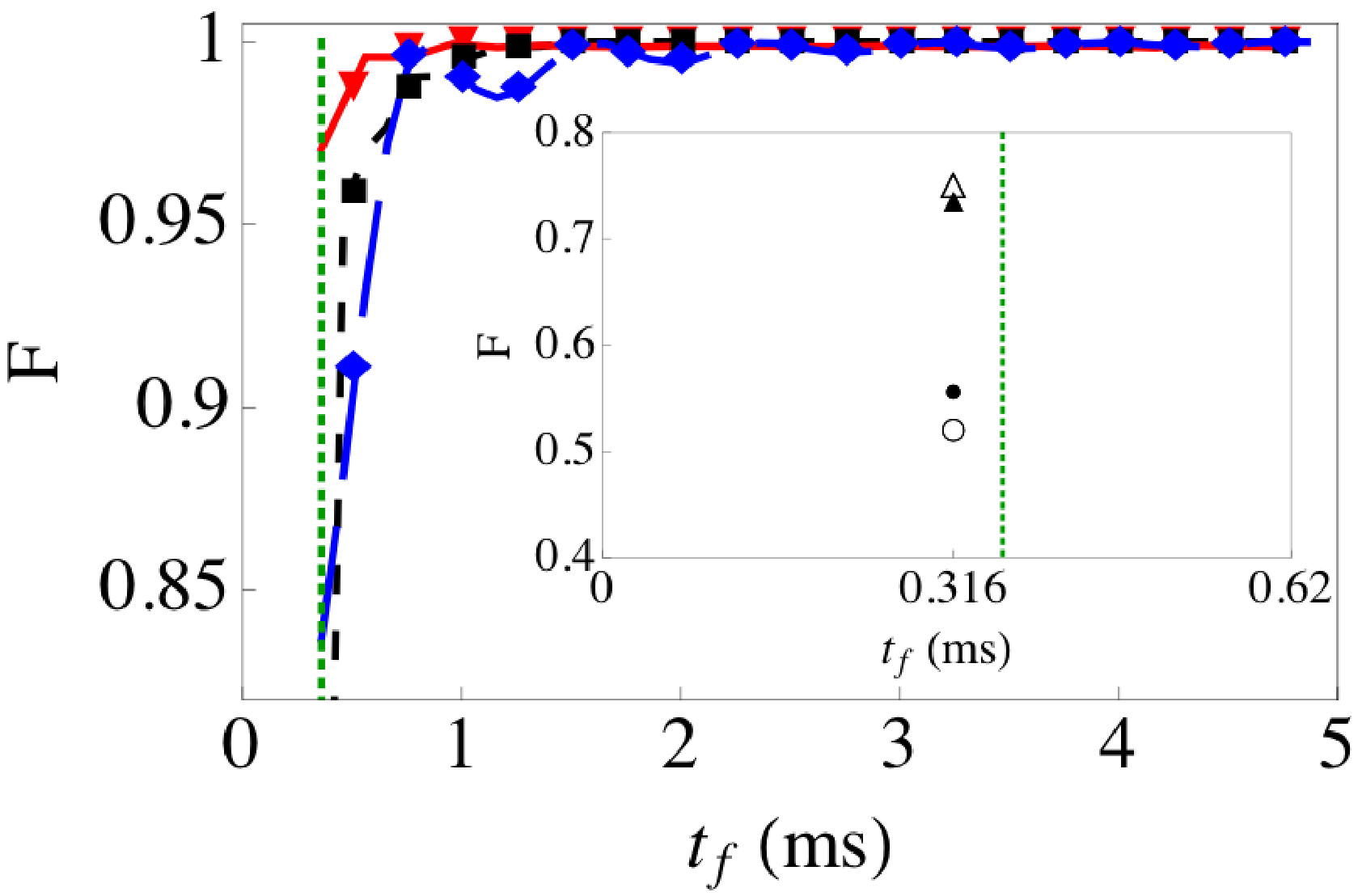}
\end{center}
\caption{\label{3d}
(Color online) Fidelity for the 3D ground state versus different final times.  
Parameters: $\nu=0$, $\omega_{0z}/2\pi=2500$ Hz,
$\lambda=1060$ nm,  
$\omega_{fz}/2\pi=250$ Hz.  
Protocols: 
inverse engineering, quintic $b$, 
$w_0=10$ $\mu m$ (red solid line), and radial fidelity (red filled triangles); 
inverse engineering, quintic $b$,  
$w_0=3$ $\mu m$ (black short-dashed line), and radial fidelity (filled squares);
bang bang, $w_0=10$ $\mu m$ (empty circle in inset), and radial fidelity (filled circle); 
bang bang, $w_0=3$ $\mu m$ (empty triangle in inset), and radial fidelity (filled triangle in inset); 
fast adiabatic, $w_0=3$ $\mu m$ and $w_0=10$ $\mu m$ (blue long-dashed line), and longitudinal fidelity (blue diamonds).
The green vertical line marks the threshold for positive trap frequencies
in the inverse engineering, quintic-$b$ protocol.    
}\end{figure}
%
%
Finally we study numerically the actual, coupled dynamics 
driven by the exact full potential, Eq. (\ref{V}),    
combining all the effects considered so far and the
longitudinal-radial coupling. The 3D fidelities are also compared with simpler 1D fidelities. 

Before discussing the results 
a remark on the technicalities of the numerical calculations of the time-dependent wave functions  
is in order. To solve the longitudinal time dependent Schr\"odinger equation 
we approximate the evolution operator  $U_L$ using the Split Operator Method (SOM) \cite{diodo10} and Fast Fourier Transform (FFT) \cite{ronnie,recipies}. 
For the radial time dependent Schr\"odinger equation,   
we have used instead the finite differences Crank-Nicholson scheme 
\cite{recipies}. In this way the singular point $r=0$ can be excluded imposing 
a zero of the wavefunction there 
for all $t$. For the 3D calculation we use similarly the Crank-Nicholson scheme with operator splitting for multidimensions (the radial and longitudinal directions) \cite{recipies}.

We calculate,   
starting from the ground state of the initial trap for $\nu=0$, the fidelity of the final state with respect to the ground state of the final trap using the three expansion protocols. 
Figure \ref{3d} clearly shows that the inverse engineering protocol provides the best 
overall performance, with a fidelity decay at smaller times that can be avoided by  increasing the waist. The radial-longitudinal quartic coupling does not 
play any significant role for the parameters of the example, 
and in any case it may also be suppressed by a waist increase.

As for the simple bang-bang approach, it fails in 3D, as expected, due to its poor radial behavior. (The radial fidelity alone is close to the 3D fidelity.)  
The results for fast adiabatic and inverse engineering methods are not too different from each other for the chosen expansion, but the fast adiabatic method will fail sooner for more demanding expansions with smaller final frequencies, as clearly illustrated in Fig. \ref{f1} (b), due to its inability to remain really adiabatic for the longitudinal direction. The main limitation of the inverse engineering  method is instead the possible failure of adiabaticity in the tighter radial direction, so it is intrinsically more robust than the fast adiabatic one. Note that the 3D fidelity is 
mimicked accurately by the 1D radial fidelity for inverse engineering and by the 1D longitudinal fidelity for the fast adiabatic approach.    
\section{Discussion and outlook\label{dao}}
In this work we have analyzed the practical implementation of 
a transitionless expansion in a simple Gaussian optical dipole trap,
taking into account anharmonicities, radial-longitudinal couplings, and the radial-longitudinal frequency mismatch. 
The main conclusion of the study is that the transitionless expansions 
in optical traps are feasible under realistic conditions. 
Despite the relation between the longitudinal and transversal trapping frequencies
through the intensity, the different timescales enable us to design fast expansions with high fidelities with respect to the ideal results using the invariant-based
inverse engineering method, which is particularly suitable compared to the two other approaches examined. Our detailed analysis of radial and longitudinal motions 
reveals the  
weakest points of each approach: 
for the inverse engineering, the main perturbation is due to the 
possible adiabaticity failure in the radial direction, which can be suppressed or 
mitigated by increasing the laser waist. This waist increase would also reduce 
smaller perturbing effects due to longitudinal anharmonicity or
radial-longitudinal coupling. 
The simple bang-bang approach fails because the time for the radial expansion is badly mismatched with respect to the ideal time, and   
the fast adiabatic method fails at short times because of the adiabaticity failure in the longitudinal direction.

Complications such as perturbations due to different noise types, and consideration of condensates, gravity effects, or 
the transient realization of imaginary trap frequencies
are left for separate works.    
Other extensions of the present analysis could involve the addition of a second laser for further control of the potential shape, or alternative traps.
Optical traps based on Bessel laser beams, for example, may be interesting 
to decouple longitudinal and radial motions. Transport protocols are also amenable to a similar 3D analysis.    
Finally, the combination of invariant-based inverse engineering with optimal control theory to optimize the frequency design is also a promising venue of research
\cite{Li10, transport3}. 
%
%
\section*{Acknowledgments}
We thank G. C. Hegerfeldt 
for useful discussions.
We acknowledge funding by the Basque Government
(Grant No. IT472-10) and Ministerio de
Ciencia e Innovaci\'on (FIS2009-12773-C02-01).
X. C.  thanks the Juan de la Cierva Programme, the National Natural Science Foundation of China (Grant Nos. 60806041 and 61176118) and the Shanghai Leading
Academic Discipline Program (Grant No. S30105);
E. T. the Basque Government (Grant No. BFI08.151); and D. G. O. the Agence National de la Recherche, the R\'egion Midi-Pyr\'en\'ees, the University Paul Sabatier (OMASYC project) and the Institut Universitaire de France.  
\appendix
%
%
%
%
%
%
%
%
%
%
%
%
%
%
%
%
\section{Perturbative treatment of longitudinal anharmonicity\label{apa}}
To quantify the effect of anharmonicity in the inverse engineering method 
let us first expand the longitudinal Hamiltonian retaining the quartic term,  
\beqa
V(z)&=&-\frac{V_0(t)}{1+z^2/z_R^2}+V_0(t)\approx V_0\frac{z^2}{z_R^2}-V_0\frac{z^4}{z_R^4}
\nonumber\\
&=&\underbrace{\frac{1}{2}m\omega_z^2(t)z^2}_{:=V_u}\underbrace{-\frac{1}{2}m\omega_z^2(t)\frac{z^4}{z_R^2}}_{:=V_1}.
\eeqa
The harmonic term is the unperturbed potential and the quartic
term the perturbation. 
A further simplification to evaluate the fidelities 
$F_L=|\la Z_n(t_f)|U_z(t_f,0)|Z_n(0)\ra|$ is to substitute the initial and final exact eigenstates $|Z_n(0)\ra$ and $|Z_n(t_f)\ra$ by corresponding eigenstates of the harmonic 
oscillator $|\psi_n(0)\ra$ and $|\psi_n(t_f)\ra$, see the expressions below. 
This is an excellent approximation for the lower eigenstates. 
In Fig. \ref{fnive}, for example, the overlap probability between exact and approximate states is  0.997 in the worst possible case ($n=5$, $w_0=3$ $\mu$m, and $\omega_{fz}=25$ Hz). 
In other words, the perturbation does not imply any significant deformation  
in the initial and final states. Its potential impact is instead at transient times where the energy may be high.       

We thus expand the overlap $\la\psi_n(t_f)|U_L(t_f,0)|\psi_n(0)\ra$
as $1+f_{n,n}^{(1)}+f_{n,n}^{(2)}+...$ 
in powers of $V_1$ to write   
\beq
{{F}}_L=\left|1+f_{n,n}^{(1)}+f_{n,n}^{(2)}+...\right|. 
\label{fide}
\eeq
The first order correction is given by 
\beqa
f_{n,n}^{(1)}&=&\frac{-i}{\hbar}\int_0^{t_f} dt'\la \psi_n(t')|V_1(t')|\psi_n(t')\ra 
\nonumber\\
&=&\frac{i m}{2\hbar z_R^2}\!\int_0^{t_f}\! dt' \omega_{L}^2(t')\la \psi_n(t')|z^4|\psi_n(t')\ra,
\label{a1}
\eeqa
and the unperturbed wavefunction corresponds to the known harmonic evolution of an expanding mode under the time-dependent frequency $\omega_z(t)$ from an eigenstate of the initial harmonic oscillator \cite{Ch10},  
\beqa
\la z|\psi_n(t')\ra&=&\frac{e^{\frac{im}{\hbar} \frac{\dot{b}z^2}{2b}}}{(b2^nn!)^{1/2}}
e^{-i(n+1/2)\omega_{0z}\int_0^{t'} \frac{dt''}{b^2(t'')}}
\nonumber\\
&\times&\left(\frac{m\omega_{0z}}{\pi\hbar}\right)^{\!\!1/4} 
e^{-\frac{m\omega_{0z} z^2}{2\hbar b^2}} H_n\left(\sqrt{\frac{m\omega_{0z}}{\hbar}}\frac{z}{b}\right).
\nonumber\\ 
\eeqa 
This is nothing but the eigenvector of the quadratic invariant times the Lewis-Riesenfeld phase factor.

Using the triangular inequality $|x+y|\geq \big||x|-|y|\big|$, ${{F}}_L\geq 1-|f_{n,n}^{(1)}+f_{n,n}^{(2)}+...|$, and
assuming that the perturbative corrections satisfy 
$|f_{n,n}^{(1)}|>>|f_{n,n}^{(2)}|$ then 
\beq
{{F}}_L\geq 1-|f_{n,n}^{(1)}|. 
\eeq
The relevant matrix element in Eq. (\ref{a1}) can be calculated explicitly,  
\beq
\la \psi_n(t')|z^4|\psi_n(t')\ra=
\left(\frac{\hbar}{m\omega_{0z}}\right)^2\frac{3b^4}{4}[(n+1)^2+n^2]. 
\eeq
Using this result and the Ermakov equation, 
\beq
|f_{n,n}^{(1)}|=\frac{3\hbar}{8 m z_R^2} [(n+1)^2+n^2]\left(t_f-\frac{1}{\omega_{0z}^2}
\int_0^{t_f}\ddot{b}b^3 dt\right). 
\eeq
To bound $-\int_0^{t_f} \ddot{b} b^3 dt$ subjected to the imposed constraints
on $b$ at $t=0$ and $t_f$
we follow closely the method applied in \cite{ChenET10} to bound the average energy. 
We integrate first by parts applying $\dot{b}(0)=\dot{b}(t_f)=0$ and minimize the resulting positive integral $\int_0^{t_f} b^2\dot{b}^2 dt$, using the Euler-Lagrange equation 
$\ddot{b}b+\dot{b}^2=0$ for $b(0)=1$, $b(t_f)=\gamma$. These two conditions are fulfilled for a set of functions larger than the subset of functions that satisfy all boundary conditions in Eq. (\ref{boco}), so the minimization sets a lower bound for the later subset.  
The solution is $b=[t(\gamma^2-1)/t_f+1]^{1/2}$ and with it the integral becomes
\beq
\label{intbound}
\int_0^{t_f} b^2 \dot{b}^2 dt=\frac{(\gamma^2-1)^2}{4 t_f}.
\eeq 
This $b$ is not as such a physically valid solution,
as the derivatives would jump at the edges,
but it maximizes the fidelity lower bound,    
\beqa
{{F}}_L&\geq& 1 -\left\{\frac{3\hbar\lambda^2}{8m\pi^2 w_0^4}[(n+1)^2+n^2]\right.
\nonumber\\
&\times&\left.\left[
t_f+\frac{3(\gamma^2-1)^2}{4 t_f \omega^2_{0z}}\right]\right\}.
\label{Fbound}
\eeqa
Take note that there is an optimal $t_f$ 
value, and that the fidelity deteriorates when increasing the  
quantum number and improves by increasing the waist. 
     
Using for $b$ the quintic polynomial in Eq. (\ref{ieb}) 
we get, instead of Eq. (\ref{intbound}),     
\beq
\int_0^{t_f} b^2 \dot{b}^2 dt= \frac{10(\gamma^2-1)^2}{24871 t_f} (1101+1351\gamma+1101\gamma^2). 
\eeq
This tends to $0.44 \gamma^4/t_f$ for $\gamma>>1$, which, up to the constant factor, is the same dependence found for the
optimal function.

The first-order analysis provides analytical results and some 
qualitative guidance but for more accurate results we should resort to a second order 
calculation. Instead of Eq. (\ref{fide}) 
we may also write the fidelity for the $n-th$ state in terms of the corresponding probability as $F_L=(1-\sum_{n\neq n'} P_{n'})^{1/2}$, where $P_{n'}$ is the probability to find the system in the $n'$-th state at $t_f$.  
This is approximated to second order using first order non-diagonal terms,     
\beq
{{F}}_L=\sqrt{1-\sum_{n\neq n'}\bigg|f_{n,n'}^{(1)}\bigg|^2},
\label{Fper1}
\eeq
where 
\beqa
f_{n,n'}^{(1)}&=&\frac{-i}{\hbar}\int_0^{t_f} dt'\la \psi_n(t')|V_1(t')|\psi_{n'}(t')\ra 
\nonumber\\
&=&\frac{i\hbar\lambda^2}{2\pi^2mw_{0}^4\omega_{0z}^{2}}\frac{\alpha_{n,n'}\beta_{n,n'}(t)}{\sqrt{\pi2^{n+n'}n!n'!}},
\label{Fper2}
\eeqa
\beq
\alpha_{n,n'}=\int_{-\infty}^{\infty} dy\ e^{-y^2} H_n(y)H_{n'}(y)y^{4},
\eeq
and
\beq
\beta_{n,n'}(t)=\int_0^{t} dt_1\ b^{4}(t_1)\omega_{z}^2(t_1)e^{-i(n'-n)\omega_{0z}\int_{0}^{t_1} \frac{dt_2}{b^2(t_2)}}.
\label{i1}
\eeq
\section{The condition of adiabaticity\label{apb}}
Consider a time-dependent Hamiltonian $\hat H_0(t)$, with instantaneous eigenstates and energies given by
\beq
\hat H_0(t) |i(t)\rangle = E_i(t)| i(t)\rangle.
\eeq
The adiabaticity condition is given by \cite{Schiff}
\beq
\label{adi_cond}
|\langle i(t)|\partial_tj(t)\rangle| \ll \frac{1}{\hbar}|E_i(t)-E_j(t)|,\quad i\neq j.
\eeq
Now we apply this general expression to get the adiabaticity condition on the longitudinal and radial frequencies
of the Gaussian beam trap. In the harmonic approximation the longitudinal Hamiltonian is given by Eq. (\ref{hz}), 
\beq
\label{harmL}
H_{har}(z,t)=-\frac{\hbar^2}{2m}\frac{\partial^2}{\partial z^2}+\frac{m\omega_{z}^{2}(t)z^2}{2},
\eeq
whereas the radial Hamiltonian is given, according to 
Eqs. (\ref{hr}), (\ref{happ}), and (\ref{omr}), by 
\beqa
H_{har}(r,t)&=&-\frac{\hbar^2}{2m}\frac{\partial^2}{\partial r^2}+\frac{m\omega_{R}^{2}(t)r^2}{2}\nonumber\\
&+&\frac{\hbar^2}{2m}\left(\frac{\nu^2-1/4}{r^2}\right). \label{harmR}
\eeqa
The instantaneous eigenstates and energies of the Hamiltonian, Eq. (\ref{harmL}), are
\beqa
\langle z|n(t)\rangle&=&\left(\frac{m\omega_z}{\pi\hbar}\right)^{1/4} \sqrt{\frac{1}{2^nn!}}
e^{-\frac{-m\omega_z z^2}{2\hbar}} H_n\left(\sqrt{\frac{m\omega_z}{\hbar}}z\right), \nonumber \\
E_{n}(t)&=&\bigg(n+\frac{1}{2}\bigg)\hbar\omega_z, \quad n=0, 1, ... \nonumber
\eeqa
where $\omega_z=\omega_z(t)$ and $H_n$ is the Hermite polynomial.
To get the adiabaticity condition for the ground state $i=0$ of the longitudinal Hamiltonian we replace these expressions for the
eigenstates and energies into Eq. (\ref{adi_cond}). The overlap of the ground state with the first excited state $j=1$ is 0 so the first non-vanishing
overlap of the ground state is with the $j=2$ state. The adiabatic condition is satisfied if
\beq
\frac{\sqrt{2}\dot\omega_z}{8\omega_z^{2}}\ll 1.
\eeq
For the radial Hamiltonian, Eq. (\ref{harmR}), the instantaneous eigenstates and energies for $\nu=0$ are \cite{REF}
\beqa
\langle r|k(t)\rangle&=&\sqrt{\frac{2m\omega_{_R}r}{\hbar}}
e^{-\frac{-m\omega_{_R} r^2}{2\hbar}} L_{k}^{0}\left(\frac{m\omega_{_R}}{\hbar}r^2\right), \nonumber \\
E_{k}(t)&=&(2k+1)\hbar\omega_{_R}, \quad k=0,  1, ... \nonumber
\eeqa
$k$ being the radial quantum number and $L_{k}^{0}$ the generalized Laguerre polynomial. Again 
$\omega_R=\omega_R(t)$. To get the adiabaticity condition for the radial direction we replace these 
expressions for $i=0$ and $j=1$ into Eq. (\ref{adi_cond}),
\beq
\frac{\dot\omega_R}{4\omega_{_R}^{2}}\ll 1.
\eeq
%
%
%
%
%
%
%
%
%
%
%

%


\begin{thebibliography}{10}
\bibitem{Phillips} A. Kastberg, W. D. Phillips, S. L. Rolston, and R. J. C. Spreeuw, 
Phys. Rev. Lett. {\bf 74} , 1542 (1995).
\bibitem{Kett}A. E. Leanhardt et al., Science {\textbf 301}, 1513 (2003). 
%
\bibitem{Hansch}W. H\"ansel, P. Hommelhoff, T. W. H\"ansch, and J. Reichel, 
Nature \textbf{413}, 498 (2001).
\bibitem{Kino} T. Kinoshita, T. Wenger, and D. S. Weiss, 
Phys. Rev. A \textbf{71}, 011602 (2005).  
\bibitem{ChenET10} X. Chen and J. G. Muga, Phys. Rev. A \textbf{82}, 053403 (2010).
\bibitem{Salamon09} P. Salamon, K. H. Hoffmann, Y. Rezek, and  R. Kosloff, Phys. Chem. Chem. Phys. \textbf{11}, 1027 (2009).
\bibitem{Muga09}J. G. Muga, X. Chen, A. Ruschhaupt, and D. Gu\'ery-Odelin, J. Phys. B \textbf{42}, 241001 (2009).
\bibitem{Ch10} X. Chen, A. Ruschhaupt, S. Schmidt, A. del Campo, D. Gu\'ery-Odelin, and J. G.  Muga, Phys. Rev. Lett. \textbf{104}, 063002 (2010).
\bibitem{Muga10} J. G. Muga, X. Chen, S. Ib\'{a}\~{n}ez, I. Lizuain, and A. Ruschhaupt, J. Phys. B \textbf{43}, 085509 (2010).
\bibitem{MN10} S. Masuda and K. Nakamura, Proc. R. Soc. A \textbf{466}, 1135 (2010).
\bibitem{Nice10} J. F. Schaff, X. L. Song, P. Vignolo, and G. Labeyrie, Phys. Rev. A \textbf{82}, 033430 (2010); \textbf{83}, 059911(E) (2011).
\bibitem{Li10} D. Stefanatos, J. Ruths, and Jr-Shin Li, Phys. Rev. A \textbf{82}, 063422 (2010).
\bibitem{MN11} S. Masuda and K. Nakamura, Physical Review A \textbf{84}, 043434 (2011). 
\bibitem{Nice11}J. F. Schaff, X. L. Song, P. Capuzzi, P. Vignolo, and G.  Labeyrie, Europhys. Lett. \textbf{93}, 23001 (2011).
\bibitem{Nice11b} J. F. Schaff, P. Capuzzi, G. Labeyrie
and P. Vignolo, arXiv:1105.2119v1.
\bibitem{optimal_control} B. Andresen, K. H. Hoffmann, J. Nulton, A. Tsirlin, and P. Salamon, Eur. J. Phys. \textbf{32}, 827 (2011).
\bibitem{Wu11}Y. Li, L.-A. Wu, and Z.-D. Wang, Phys. Rev. A \textbf{83}, 043804 (2011).
\bibitem{Adol11} A. del Campo, Phys. Rev. A \textbf{84}, 031606(R) (2011).
\bibitem{Adol11b} A. del Campo, Eur. Phys. Lett. (accepted), arXiv:1010.2854.
\bibitem{Chen11} X. Chen, E. Torrontegui, and J. G. Muga, Phys. Rev. A \textbf{83}, 062116 (2011).
\bibitem{nonHermitian} S. Ib\'{a}\~{n}ez, S. Mart\'{i}nez-Garaot, X. Chen, E. Torrontegui, and J. G. Muga, Phys. Rev. A \textbf{84}, 023415 (2011).
\bibitem{Onofrio}S. Choi, R. Onofrio, and B. Sundaram, Phys. Rev. A (accepted), arXiv:1109.4908.
\bibitem{odt} R. Grimm, M. Weidem\"uller, and T. B. Ovchinnikov,  Adv. At. Mol. Opt. Phys. \textbf{42}, 95 (2000). 
\bibitem{Vogels} J. M. Vogels, personal communication. 
\bibitem{transport3} X. Chen, E. Torrontegui, D. Stefanatos, Jr-Shin Li, and J. G. Muga, Phys. Rev. A {\textbf{84}}, 043415 (2011). 
\bibitem{Nemoto} S. Nemoto, Appl. Opt. \textbf{29}, 1940 (1990).
\bibitem{Schiff} L. I. Schiff, {\it Quantum Mechanics} (McGraw Hill, New York, 1949). 
\bibitem{Rice03}M. Demirplak and S. A. Rice, J. Phys. Chem. \textbf{107}, 9937 (2003).
\bibitem{diodo10}
E. Torrontegui, J. Echanobe, A. Ruschhaupt, D. Gu\'ery-Odelin, and J. G. Muga, 
Phys. Rev. A. \textbf{82}, 043420 (2010).
\bibitem{ronnie} 
R. Kosloff, J. Phys. Chem \textbf{92}, 2087 (1988). 
\bibitem{recipies}
W. H. Press, S. A. Teukolsky, W. T. Vetterling, and B. P. Flanery, {\it Numerical Recipies The Art of Scientific Computing $3$rd ed.} (Cambridge University Press, Cambridge, 2007).
\bibitem{REF} P. Camiz A. Gerardi, C. Marchioro, E. Presutti, and E. Scacciatelli, J.  Math. Phys. \textbf{12}, 2040 (1971).






%

\end{thebibliography}
\end{document}